\documentclass{sig-alternate-2013}

\usepackage{graphicx}
\usepackage{enumitem}
\usepackage{theorem}
\usepackage{hyperref}
\usepackage{times}
\usepackage{url}
\usepackage{color} 
\usepackage{comment}
\usepackage[labelfont=bf,labelsep=period,justification=raggedright]{caption}

\theoremstyle{remark}
 \newtheorem{mydef}{Definition}

\newcommand{\junk}[1]{}
 \permission{Copyright is held by the International World Wide Web Conference Committee (IW3C2). IW3C2 reserves the right to provide a hyperlink to the author's site if the Material is used in electronic media.}
\conferenceinfo{WWW'14 Companion,}{April 7--11, 2014, Seoul, Korea.} 
\copyrightetc{ACM \the\acmcopyr}
\crdata{978-1-4503-2745-9/14/04. \\
http://dx.doi.org/10.1145/2567948.2579038}

\begin{document}

\title{People Like Us: Mining Scholarly Data for Comparable Researchers}
      
%
%
%
%
%

\numberofauthors{3} 
%
\author{
%
%
\alignauthor
Graham Cormode\\
       \affaddr{University of Warwick}\\
          \email{G.Cormode@warwick.ac.uk}
\alignauthor
S. Muthukrishnan\\
       \affaddr{Rutgers University}\\
        \email{muthu@cs.rutgers.edu}
\alignauthor Jinyun Yan\thanks{Authors by alphabetical order} \\
       \affaddr{Rutgers University}\\
        \email{jinyuny@cs.rutgers.edu}
}
 

\maketitle

\begin{abstract}
We present the problem of finding comparable researchers for any given researcher. 
This problem has many motivations.
Firstly, know thyself.
The answers of where we stand among research community and who we are
most alike may not be easily found by existing evaluations of ones' research 
mainly based on citation counts.
Secondly, there are many situations where one needs to find comparable researchers e.g., for reviewing peers, constructing programming committees or
compiling teams for grants. 
It is often done through an ad hoc and informal basis. 

Utilizing the large scale scholarly data accessible on the web, we address the problem of 
automatically finding comparable researchers.
We propose a standard to quantify the quality of research output, via the quality of publishing venues. 
We represent a researcher as a sequence of her publication records, and develop a framework of comparison of researchers by sequence matching. 
Several variations of comparisons are considered including matching by
quality of publication venue and research topics, and performing prefix matching.  
We evaluate our methods on a large corpus and
demonstrate the effectiveness  of our methods through examples.
In the end, we identify several promising directions for further work. 
\end{abstract}

\vspace{1mm}
\noindent
{\bf Categories and Subject Descriptors:} \\ 
I.2.6 [{\bf Artificial Intelligence}]: {Learning}

 \vspace{1mm}
 \noindent
 {\bf General Terms:} Algorithms, Experimentation, Measurement

 \vspace{1mm}

\allowdisplaybreaks

\section{Introduction} 

\junk{
In research, it is a common task to look for comparable people. 
Recommendation letters and tenure cases often suggest other
researchers who are comparable to the individual in question. 
The situation is similar in finding peer reviewers, journal editors and programming committees for conferences.
Yet finding the right person to compare against is a
vexing task. 
There is no simple strategy that  allows a similar person to be
found. 
Natural first attempts, such as looking at co-authors, or scouring the
author's preferred publication venues, either fail to find good
candidates, or swamp us with too many possibilities.

There are limited number existing metrics to evaluate research impact at the individual level.
Examples include h-index ~\cite{hirsch2005index} and g-index~\cite{egghe2006improvement}.
These metrics are mainly based on raw citation counts, i.e., the number of papers citing a given paper, which have several limitations. 
Firstly, Garfield~\cite{garfield1972citation} argued that citation
counts are a function of many influencing factors besides scientific
quality, such as area of research, 
number of co-authors
the language used in the paper
and various other factors. 
Secondly, in many cases few if any citations are recorded, even though
the paper's influence may go beyond this crude measure of impact~\cite{meho2006rise}. 
Thirdly, citation counts evolve over time.  Papers published longer
ago are more likely to have higher counts than those released more
recently. 

\smallskip
\noindent
{\bf  Our Approach.}
Focusing on the computer science domain, we propose an approach to
compare researchers that utilizes the quality of venues of 
publication. 
In this paper, we focus on conferences, since researchers in computer science often prefer conference publications, and the data available on the web is also skewed to conferences. 
Other disciplines may favor journals instead; 
our methods apply equally to such settings. 

While citation behavior varies across sub-fields,
we can treat the quality rank of venues as a way to level the
comparison across sub-fields. 
We associate a paper with the quality rank of its publishing
venue.\footnote{We adopt this as a convenient shorthand for the quality
  of the paper; alternate methods for assigning a quality to a paper
  can also be used here.} 
Instead of averaging over the quality rank, which might be unreliable in comparison, we consider the sequence of venue qualities over the full career of a researcher.   
 
Our key intuition is that the career trajectory of a researcher can be
represented as a series of their publications. 
We use the quality of the venue as a surrogate for the quality of the
paper.  
Consequently we can compare two researchers by matching their career
trajectories, as sequences of venue rankings. 
The distance between two researchers is calculated by allowing some
mismatches, and counting the number of deletion and insertion operations
necessary to harmonize the two sequences. 
Besides the pattern on the quality of publishing venues, we also
consider research topics to identify 
comparable researchers in the same or similar sub-fields.  
We thus propose a variant that incorporates  topic similarity
between authors. 
With simple modifications our methods can be used to match a junior
author to the early career stages of a more senior researcher. 
This can be especially useful when trying to predict the trajectory of
a researcher for years to come. 
 
 \medskip
 \noindent
 {\bf Data.}
 There are many online services that index research work. 
For computer science, the
DBLP\footnote{\url{http://www.informatik.uni-trier.de/~ley/db/}} is a
bibliography website that lists more than 2.3 million articles; while
arXiv.org~\footnote{\url{http://arxiv.org/}} hosts hundreds of
thousands of pre-prints from computer science and beyond. 
Services such as Google Scholar\footnote{\url{scholar.google.com}},  arnetMiner\footnote{\url{http://arnetminer.org/}}, researchGate\footnote{\url{http://www.researchgate.net/}} offer rich functionalities including search, information aggregation and navigation, and social networking.
The availability of such data has led to its use for numerous other
applications. 
For example, metrics
such as h-index~\cite{hirsch2005index} to evaluate the impact of a
researcher;  the network structure of scientists connected by
co-authorship relation~\cite{newman2004coauthorship}; community
detection in citation networks~\cite{leskovec2010empirical}; the study
of how science is written~\cite{cormode2012scienceography}.  

Other services provide rankings of publication venues, e.g. Google Scholar
Metrics\footnote{\url{http://scholar.google.com/intl/en/scholar/metrics.html}},
Microsoft
Academic~\footnote{\url{http://academic.research.microsoft.com/}}and
CORE\footnote{\url{http://core.edu.au/index.php/categories/conference\%20rankings}}. 
While coverage of venues is large we found there is considerable
disagreement among sources in categorizing sub-fields, 
and many ranking results may appear surprising. 
How to rank topic-dependent venues objectively remains an interesting
and open research problem. 
To simplify the process and focus on the comparison algorithms, we take advantage of an existing subject-dependent ranking that covers broadly known conferences. 

 \junk{
In this paper, our contributions are 1) finding an objective standard to evaluate individual's research output and to compare pairs of researches; 2) defining comparable relation and proposing practical methods to determine such relation; 3) multiple scenarios are considered: comparable relation by full matching on publication quality, on both topic similarity and publication quality, and prefix matching; 4) conducting exploratory analysis and experiments on large scale datasets, which contain millions of researchers and publications.

\subsection{On Topic Similarity}

Although our corpus is focused on computer scientists,
the computer science discipline spans a range of topics from theoretical studies of algorithms to computing systems in hardware and software.
For real world applications, it is more common to compare researchers
who work on the same or similar areas. 
When the pool of candidates is filtered before the evaluation and comparison, our method can be directly applied. 
If such prior information is not available, we propose to learn topic interests of researchers then compare them automatically. 
In other words, we can detect both {\em similar} and {\em comparable} people simultaneously. 
Our main intuition is that the matching distance of two points in two sequences can depend on both venue score and topic similarity. 

We design a new distance metric to integrate both topic similarity and venue quality. 
Given two sequences $S$ and $R$ corresponding to authors $a_S$ and $a_R$,  
the matching between $i$-th point in $S$ and $j$-th point in $R$ is calculated as
\[
	d(s_i, r_j) = | v_i - v_j + \epsilon | \cdot w(p_i, p_j ) 
\]
where $v_i$  and $p_i$ are $i$-th venue score and paper for author
$a_S$;  $v_j$ and $p_j$ are the corresponding venue score and paper
for  author $a_R$ (with venue score as an ordinal variable); and
$\epsilon$ is a small constant, discussed below. 
The topic distance $w(p_i, p_j)$ depends on the topic similarity of two papers $p_i$ and $p_j$, and is computed as $ w(p_i, p_j) =  1- \operatorname{sim}(p_i, p_j)$.
The value of topic similarity of two papers $\operatorname{sim}(p_i,
p_j)$ is in [0,1]. 
If two papers are on similar topics, the topic distance is small,
otherwise it is big.

When venue scores $v_i, v_j$ are the same, in the previous algorithm,
the distance is zero. However the topic distance might be large.
We introduce the constant $\epsilon$ to include the topic distance for
points with same venue score. 
In our experiments, $\epsilon$ is set to be $0.1$.
Based on the above definition, we see if the topic distance is small and the venue score distance is small, the distance between these two points is small. 

The topic similarity between two papers is computed based on the
content of papers. 
In our corpus, we have the title for all papers,  
and abstracts for about a third of papers. 
To discover topics for papers,  we implement Latent Dirichlet Allocation
(LDA)~\cite{hoffman2010online}. 
We treat the concatenation of title and abstract of a paper as a
document. Topics are derived from the whole corpus. We then obtain
the topic distribution for each paper. 
The main parameter is the number of topics.  
We experimented with 20, 50 and 100 topics, with manual validation on frequent words in each topic, and select the number of topics which provides the best presentation of topics. 

\begin{table*}[t]
\caption{Case Study of Topic Edit Distance}
\vspace*{-0.1in}
\centering
\begin{small}
\begin{tabular}{|p{1.6cm}|p{2cm}|p{13cm} |}
\hline
Topic & Researcher &  Top 20 Comparable Researchers  \\
\hline
Theory & \href{http://en.wikipedia.org/wiki/Richard_M._Karp}{Richard M. Karp}  & 
\href{http://en.wikipedia.org/wiki/David_Karger}{David R. Karger}, 
Ravi Kumar, 
\href{http://en.wikipedia.org/wiki/Jeffrey_Ullman}{Jeffrey D. Ullman}, 
\href{http://en.wikipedia.org/wiki/Avrim_Blum}{Avrim Blum}, 
\href{http://en.wikipedia.org/wiki/Joseph_Seffi_Naor}{Joseph Naor},
\href{http://en.wikipedia.org/wiki/F._T._Leighton}{Frank Thomson Leighton}, 
\href{http://en.wikipedia.org/wiki/Rajeev_Motwani}{Rajeev Motwani}, 
\href{http://en.wikipedia.org/wiki/Hari_Balakrishnan}{Hari Balakrishnan}, 
\href{http://en.wikipedia.org/wiki/Eric_Horvitz}{Eric Horvitz},
Mostafa H. Ammar, 
Rina Dechter, 
\href{http://en.wikipedia.org/wiki/Prabhakar_Raghavan}{Prabhakar Raghavan}, 
Craig Boutilier, 
\href{http://en.wikipedia.org/wiki/Rafail_Ostrovsky}{Rafail Ostrovsky}, 
\href{http://en.wikipedia.org/wiki/Raghu_Ramakrishnan}{Raghu Ramakrishnan},
 Yossi Azar, 
\href{http://en.wikipedia.org/wiki/James_Kurose}{James F. Kurose}, 
\href{http://en.wikipedia.org/wiki/Josep_Torrellas}{Josep Torrellas},
\href{http://en.wikipedia.org/wiki/Rakesh_Agrawal_(computer_scientist)}{Rakesh Agrawal}, 
\href{http://en.wikipedia.org/wiki/Andrew_Ng}{Andrew Y. Ng}
\\
\hline
Machine Learning & Judea Pearl  & Craig Boutilier, Satinder P. Singh, Avrim Blum, Manfred K. Warmuth, Michael J. Kearns, Piotr Indyk, Eyal Kushilevitz, Surajit Chaudhuri, Yoram Singer, Robert E. Schapire, Jon M. Kleinberg, Shafi Goldwasser, Robert Endre Tarjan, Geoffrey E. Hinton, Eric Horvitz, Milind Tambe, Jeffrey S. Rosenschein, Silvio Micali, Daniel S. Weld, Nick Koudas     \\
\hline
 Networks & Hari Balakrishnan & 
Ion Stoica, James F. Kurose, Baochun Li, Gustavo Alonso, Mostafa H. Ammar,
Eitan Altman, Robert Endre Tarjan, Surajit Chaudhuri, Jon M. Kleinberg, Ness
B. Shroff, Yossi Azar, Eli Upfal, Peter Steenkiste, Joseph Naor, Sang Hyuk
Son, Qian Zhang, Frank Thomson Leighton, Randy H. Katz, Hagit Attiya,
Wang-Chien Lee\\
\hline
Distributed Computing & Nancy Lynch &
Baruch Awerbuch, Scott Shenker, J. J. Garcia-Luna-Aceves, Sajal K. Das,
Roger Wattenhofer, Moni Naor, Hossam S. Hassanein, Rachid Guerraoui, Amr El
Abbadi, David E. Culler, Yishay Mansour, Christos H. Papadimitriou, Klara
Nahrstedt, Danny Dolev, Christos Faloutsos, Deborah Estrin, Mostafa H.
Ammar, Mario Gerla, Lionel M. Ni, Serge Abiteboul\\
\hline 
Computer Vision & Dimitris N. Metaxas   &  Andrew Blake, Trevor Darrell, Jitendra Malik, Jean Ponce, Narendra Ahuja, Shaogang Gong, Dale Schuurmans, Stefano Soatto, Alan L. Yuille, Pascal Fua, Aly A. Farag, Pedro Domingos, Shree K. Nayar, Xilin Chen, Chris H. Q. Ding, Brendan J. Frey, Pietro Perona, Santosh Vempala, Thomas G. Dietterich, Nassir Navab  \\
\hline
Secrity & 
Dan Boneh & 
Amit Sahai, Ueli M. Maurer, Jacques Stern, Ronald L. Rivest, Ran Canetti,
Shafi Goldwasser, Stuart J. Russell, Abraham Silberschatz, Matthew Andrews,
George Varghese, Russell Impagliazzo, Cynthia Dwork, David Heckerman, Hector
J. Levesque, Eyal Kushilevitz, Michael J. Kearns, Robert E. Schapire, Joe
Kilian, Anoop Gupta, Tatsuaki Okamoto   \\
\hline
Algorithms & Donald E. Knuth & Mark Roberts, H. Ramesh, Fady Alajaji, Hemant Kanakia, Mary E. S. Loomis, Michael D. Grossberg, Antonio Piccolboni, Paul Milgram, Andy Boucher, Paul Hart, Kazuo Sumita, Nicholas Carriero, Shiro Ikeda, James M. Stichnoth, Michael Bedford Taylor, Kimiko Ryokai, Riccardo Melen, Jatin Chhugani, Dianne P. O'Leary, Lal George  \\
\hline
\end{tabular}
\end{small}
\label{case-study2}
\end{table*}%

Given the topic distribution for each paper, we can compute the topic
similarity via cosine similarity, or Jensen-Shannon divergence, etc. 
We use cosine similarity in our examples. 
With the new distance metric, the dynamic programming formula is modified to the following.
\[
	D(i, j) =  \min
	\begin{cases}
	D(i-1, j-1) + d(s_i, r_j)  & \text{ match or mismatch }   \\
	D(i-1, j) + v(s_i),  & \text{ insertion}  \\
	D(i, j-1) + v(r_j), & \text{ deletion }  
	\end{cases}
\]
We present some examples of researchers from different topics in our
results in Table~\ref{case-study2}. 
In general, we found results for each researcher are  closer in
research topics. 
See, for example, the output for ``Dimitris N. Metaxas'' 
in Tables~\ref{case-study1} and~\ref{case-study2}. 
For ``Judea Pearl'', both edit distance and topic edit distance return comparable authors in similar research topics. 
Recall that the topic distribution of each author is learned mostly from their paper titles. 
We manually validated many examples, and compared the results by
simply utilizing the topic similarity between authors and by our
approach.  
We found that sequence matching combining topics from title and venue scores did a better job in finding authors in similar research area.
Taking ``Richard M. Karp'' for example, we find that 16 of the 20 comparable
researchers returned also have entries in Wikipedia, a crude
indication that they are similarly notable. 
Future work may more systematically examine the performance of clustering similar authors by our distance metric.  

There are a few notable bad examples: 
the comparable researchers for ``Donald E. Knuth'' are only loosely
related.  
Knuth's paper titles are often short, and commonly use
generic computer science terms like ``Algorithm''. 
Hence, topic inference on his papers has poor performance, and the comparable authors are mainly determined on venue score sequence matching.
As our data contains only $30,000$ authors, 
many are missing (along with their papers), limiting the set of
potential comparable authors. 

\subsection{On Prefix Matching}
Each year, many junior researchers begin their career.
It is useful and interesting to matching junior researchers to segments of senior researchers. 
With simple modification, our algorithm can be used to compare a
junior researcher to senior researchers in their early career stage. 
This can be useful, for example, to committees considering the future
prospects of job candidates, and to junior researchers finding out whose career trajectory they are following.  

Formally, we are interested in the problem that, given a senior
researcher and a junior researcher characterized  by $S$ and $R$ respectively. 
Instead of matching full sequence of $S$ and $R$, we want to match $R$ to every prefix of $S$. A prefix of the sequence $S$ with length $n$ is denoted by $S[1:k]$ where $1 \leq k \leq n$. The final distance is then the minimal of matching distances with every prefix. 
If we store all intermediate steps of the dynamic programming
table, we can easily compute the distance of prefix matching. Specifically, the vector $D(:,m)$ stores the minimal distance from every prefix of $S$ to $R$, where $m$ is the length of $R$. Consequently the minimal distance is no longer $D(n,m)$ but $\min D(:,m)$.  

We sample authors with fewer than 100 papers within less than 20 years
of research period to test the prefix matching.  
For brevity, we omit examples. 
Comparing to matching full sequence, there are more senior researchers mixed in the results by prefix matching.

\section{Conclusions and Future Work}
\label{sec:conclusion}
In this paper, we address the novel problem of automatically finding
comparable researchers through large scholarly data.
Unlike existing work, which evaluates researchers mainly by citation
counts, our methods consider the sequence of the quality of publishing
venues, which seems more appropriate for evaluating and comparing
research output. To allow automatic identification of comparable
people in similar research areas, we further propose a distance metric which
combines the topic similarity and venue quality.  
Our approach can be easily modified to match junior researches to senior researcher at their beginning of research periods.
 
Our analysis and experiment was conducted on large-scale scholarly datasets available on the web. The effectiveness of our methods are demonstrated by arbitrarily picked examples. 
There are several problems open for future study.
\begin{itemize}[noitemsep,nolistsep]
\item Data Collection: Lack of data may lead to less accurate
  results. 
Many challenges exist in the data collection, e.g. reducing the language gap, 
knowledge extraction from multiple data sources with different formats.
   \item Evaluation:  
  There is currently no ``ground truth'' for our methods. 
  We are developing a user interface to allow exploration of
  comparable people, and collect user feedback on results. 
  \item Optimization: Our methods compute the matching distance
    between each pair of researchers through their full publication
    records, a quadratic number of comparisons. 
  A different approach is required to make this more scalable. 
  \item Comparable Network: With comparable relation established, we
    can define a comparable network, in which each node is a
    researcher, and edges connect comparable nodes.
The weight on the edge is related to the distance between two nodes. It may be interesting to examine the structure of such network, and compare it with co-authorship and citation networks.
\end{itemize}

\section{Acknowledgement}
This work was sponsored by the NSF Grant 1161151: AF: Sparse Approximation: Theory and Extensions. 

%
\vspace*{-0.1in}
{

} 
%
%
\end{document}